\documentclass[aps,superscriptaddress,twocolumn]{revtex4}
\usepackage{amsfonts,amssymb,amsmath}            
\usepackage[]{graphics,graphicx,epsfig}            
\usepackage{amsthm}
\def\identity{\leavevmode\hbox{\small1\kern-3.8pt\normalsize1}}

\newcommand{\ket}[1]{\left | #1 \right\rangle}
\newcommand{\bra}[1]{\left \langle #1 \right |}
\newcommand{\half}{\mbox{$\textstyle \frac{1}{2}$}}

\newcommand{\braket}[2]{\left\langle #1|#2\right\rangle}
\newcommand{\proj}[1]{\ket{#1}\bra{#1}}
\renewcommand{\epsilon}{\varepsilon}
\begin{document}

\title{The Non-Equilibrium Reliability of Quantum Memories}
\date{\today}

\author{Alastair \surname{Kay}}
\affiliation{Max-Planck-Institut f\"ur Quantenoptik, Hans-Kopfermann-Str.\ 1,
D-85748 Garching, Germany}
\affiliation{Centre for Quantum Computation,
             DAMTP,
             Centre for Mathematical Sciences,
             University of Cambridge,
             Wilberforce Road,
             Cambridge CB3 0WA, UK}
\affiliation{Centre for Quantum Technologies, National University of Singapore, 3 Science Drive 2, Singapore 117543}

\begin{abstract}
The ability to store quantum information without recourse to constant feedback processes would yield a significant advantage for future implementations of quantum information processing. In this paper, limitations of the prototypical model, the Toric code in two dimensions, are elucidated along with a sufficient condition for overcoming these limitations. Specifically, the interplay between Hamiltonian perturbations and dynamically occurring noise is considered as a system in its ground state is brought into contact with a thermal reservoir. This proves that when utilizing the Toric code on $N^2$ qubits in a 2D lattice as a quantum memory, the information cannot be stored for a time $O(N)$. In contrast, the 2D Ising model protects classical information against the described noise model for exponentially long times. The results also have implications for the robustness of braiding operations in topological quantum computation.
\end{abstract}

\maketitle

{\em Introduction:} An essential element to the rapidly evolving technologies involved in the implementation of quantum information processing will be the ability to store quantum states in the presence of noise. The notion of {\em reliable storage} of a qubit requires that there should be a family of storage solutions parameterised by their size, $N$, and that the time for which the information survives should grow exponentially in $N$. That such solutions exist is guaranteed by the theory of fault tolerance \cite{gottesman:2005}, which states that provided errors occur locally, and only at an error rate below a threshold rate which is independent of $N$, suitable encodings of the data can be found, although regular error correction cycles are required, with massive overheads.

Evidently, a device more akin to a quantum hard drive is desirable, where the data can be input and later read-out without intervention in the meantime. Without any error correction whatsoever, this idea is clearly flawed; any state $\ket{\Psi}$ of $N$ qubits can be converted into an orthogonal state by a local rotation on any single qubit. To see this, consider the Schmidt decomposition of $\ket{\Psi}$ on a particular qubit $i$, $\ket{\Psi}=\alpha\ket{0}_i\ket{\psi_0}+\beta\ket{1}_i\ket{\psi_1}$, where $|\alpha|^2+|\beta|^2=1$ and $\braket{\psi_0}{\psi_1}=0$. A bit-flip, $X$, on this qubit has the effect $\bra{\Psi}X_i\ket{\Psi}=0$. For a noise model with a constant per-qubit error rate, the length of time for which $\ket{\Psi}$ survives decreases with $N$.

The intermediate regime involves encoding a qubit state within a subspace (or subsystem) of $N$ qubits, and allowing a single round of error correction when that state is read-out. Reliable storage remains feasible since the logical gate operations $X_L$ and $Z_L$ which act on these states, $\ket{0_L}$ and $\ket{1_L}$, can be non-local, so that local rotations do not mix the subspace, and can thus be detected and corrected at read-out. One concept to ensure that these local errors do not build up is to create a Hamiltonian in which the states $\ket{0_L}$ and $\ket{1_L}$ are degenerate ground states. If this Hamiltonian has an energy gap $\Delta$ up to the next excited state, then one might hope that for any temperature $k_BT\ll\Delta$, thermal noise would be unable to mix the ground states. This motivated Kitaev \cite{DKLP02a} to introduce the Toric code in two dimensions.

Following the introduction of a Hamiltonian $H$, one must consider both noise due to interaction with an environment and also perturbations in $H$. In this paper, we consider the interaction between these two types of error to prove that the Toric code in 2D does not provide reliable storage of information, and include the necessary details for the generalisation of the technique to other stabilizer Hamiltonians. These reveal a sufficient condition (for both stabilizer and non-stabilizer Hamiltonians) such that the presented test is passed, which requires models to exhibit a property known as string tension. The 2D Ising model displays this for classical data, as does the Toric code in 4D for quantum data \cite{DKLP02a}.

{\em The Toric Code} in two dimensions is the paradigmatic construction of a quantum memory. Consider a plane with orthogonal vectors $\hat x$ and $\hat z$ and periodic boundary conditions. Qubits are located at positions $2i\hat x+2j\hat z$ and $(2i+1)\hat x+(2j+1)\hat z$ for all integers $0\leq i,j\leq N-1$, and they interact via a Hamiltonian
$$
H=-\frac{\Delta}{2}\sum_{i,j}\bar Z_{i,j}+\bar X_{i,j},
$$
where $\bar Z_{i,j}=Z_{2i,2j}Z_{2i+1,2j+1}Z_{2i+2,2j}Z_{2i+1,2j-1}$, $\bar X_{i,j}=X_{2i,2j}X_{2i+1,2j+1}X_{2i-1,2j+1}X_{2i,2j+2}$ and $X_{i,j}$ is the standard Pauli $X$-operator acting on the qubit at position $i\hat x+j\hat z$. The ground state is 4-fold degenerate, encoding two logical qubits, and the loops $Z_{L,1}=\prod_iZ_{2i+1,2j+1}$ and $Z_{L,2}=\prod_jZ_{2i,2j}$ define the $Z$-basis of the two qubits they encode. In order to convert between these states, a string of $X$ operators such as $\prod_{i=0}^{N-1}X_{2i,2j}$ needs to be applied. It is typically argued \cite{DKLP02a} that any perturbation
$
\delta H=\delta\sum_{i}h_i,
$
where $h_i$ acts on a constant number of qubits in a local way, and satisfies $\|h_i\|\leq 1$, only splits the degeneracy by an amount $\delta^N$, and thus information can survive for a time $\delta^{-N}$. This is because a product of $O(N)$ of the $h_i$ is required to describe a logical rotation within the degenerate subspace, and hence $O(N)^{th}$ order perturbation theory is required \footnote{Strictly, perturbation theory requires $\|\delta H\|\leq 1$, i.e.~$\delta\sim O(1/N)$, not $\delta\sim O(1)$, but any discrepancy is irrelevant in what follows; any perturbations that we introduce do not affect the ground state space or energy. This also avoids questions relating to the influence of perturbations on the ground state space.}. Evidence is mounting \cite{horodecki,nussinov-2007,nussinov-2007a,CC07,pacman} that the Toric code in 2D is not protected in the same, exponential, way against interactions with a thermal environment, a property known as self-correction. It is this which we wish to prove. The noise models that have so far been explicitly discussed have been based on a Monte-Carlo approach \cite{horodecki} which has the advantages that it is easily analyzed, and is known to have the thermal state as its steady state. However, there is no known physical process that results in this type of noise. Alternative treatments mostly consider whether storage is possible when the system is in equilibrium with a thermal bath \cite{horodecki,nussinov-2007,nussinov-2007a,CC07,pacman}, but are beginning to consider the time required to approach equilibrium from an initial state \cite{nussinov-2007a}. All these approaches take it as given that Hamiltonian perturbations are not destructive.

In order to discuss how Hamiltonian perturbations interact with errors that enter the system, we impose the same demands as required for the derivation of fault-tolerance -- robustness of storage must be with respect to an adversarial model i.e.~{\em all} possible error combinations subject to some physically motivated restrictions. By selecting one particular example, an upper bound on the protection given in the worst-case can be derived. We avoid a full treatment of an arbitrary environment by considering a minimal degree of interaction of such an environment as introducing a single fault $X_{0,0}$. Without loss of generality, the system is considered to be initialised in a ground state $\ket{\psi}$ that is the $+1$ eigenstate of both $Z_{L,1}$ and $Z_{L,2}$. The aim is to use the Hamiltonian perturbations to propagate the error through a sequence of intermediate operations $U_l=\prod_{i=0}^lX_{2i,0}$ until $U_{N-2}$ has been implemented, flipping the logical qubit, up to a single local error \footnote{The remaining error would be removed by error correction when the memory is finally read.}. If a Hamiltonian achieves this transport in a time $t$, then data cannot be stored reliably for longer times. We proceed as in \cite{kay-2006b} by defining a subspace of interest, spanned by the basis $\{U_l\ket{\psi}\}$ for $i=1\ldots N-2$. The Hamiltonian for this subspace is described by an $(N-2)\times(N-2)$ diagonal matrix with equal diagonal elements (equal to the energy of a single excitation). There are two perturbations that we can add that preserve this subspace structure. The first allows alteration of individual diagonal terms, while the second is a propagation operator, creating off-diagonal matrix elements,
\begin{eqnarray}
\half B_i(\identity-\bar Z_{i})U_l\ket{\psi}&=&\delta_{il}B_iU_l\ket{\psi}   \nonumber\\
X_{2i+2,0}(\identity-\bar Z_{i,0}\bar Z_{i+1,0})U_l\ket{\psi}&=&\left\{\begin{array}{cc}
U_{l+1}\ket{\psi} & i=l \\
U_{l-1}\ket{\psi} & i=l+1\\
0 & \text{otherwise}
\end{array}\right.. \nonumber
\end{eqnarray}
These terms combine to give a perturbation
\begin{eqnarray}
\delta H&=&\half\delta\sum_{i=0}^{N-3}J_i(X_{2{i+1},0}(\identity-\bar Z_{i,0}\bar Z_{i+1,0}))    \nonumber\\
&&+\half\delta\sum_{i=0}^{N-2}B_i(\identity-\bar Z_{i,0}).   \nonumber
\end{eqnarray}
The subspace of $H+\delta H$ containing the evolution can be written as an effective Hamiltonian $H_{\text{eff}}$, which is a tridiagonal matrix with matrix elements $\bra{i}H_{\text{eff}}\ket{i}=\bra{\psi}U_i^{\dagger}HU_i\ket{\psi}=2\Delta+\delta B_i$ and $\bra{i}H_{\text{eff}}\ket{i+1}=\delta J_i$. Matrices of this form have been heavily studied for their ability to perfectly evolve a state $\ket{i}$ into $\ket{N-2-i}$. For example, selecting $B_i=0,J_i=\frac{2}{N-1}\sqrt{(i+1)(N-2-i)}$ reproduces the scheme used in \cite{christandl}, which invokes a mapping to the $J_x$ rotation matrix of a spin $\half(N-2)$ particle. In the Toric code model, the transfer $\ket{0}$ to $\ket{N-2}$ corresponds to the transfer $U_0\ket{\psi}$ (the ground state with a single error) to $U_{N-2}\ket{\psi}$ (the ground state with a logical rotation and a single error). Thus, in a time $\sim N/\delta$ an adversarial error model is guaranteed to propagate a single error into a logical gate operation and data stored using the Toric code does not reliably maintain its integrity over longer time scales.

The more widely applicable technique of Karbach and Stolze \cite{transfer_comment} can be introduced to select the values of $J_i$ and $B_i$ serves to motivate the generality of the approach. This starts from the $M\times M$ effective Hamiltonian ($M\sim\text{poly}(N)$), which, assuming some valid assignment for the $J_i$ and $B_i$, can be diagonalized to find the eigenvalues $\lambda_i$ and eigenvectors $\ket{\lambda_i}$. The fidelity of state transfer is given by
$$
F=\left|\sum_{i=1}^Me^{-i\lambda_it}a_i\right|,
$$
where $a_i=\braket{M}{\lambda_i}\braket{\lambda_i}{1}$. The maximum achievable value, for fixed $\{a_i\}$, is $F_{\text{max}}\leq\sum_i|a_i|$. Moreover, for symmetric tridiagonal Hamiltonians ($J_i=J_{M+1-i}\neq0$), $\braket{M}{\lambda_i}=(-1)^{i}\braket{\lambda_i}{1}$, so $F_{\text{max}}=\sum_i|\braket{\lambda_i}{1}|^2=1$. By selecting new eigenvalues such that, for some $t$ and $\theta$,
\begin{equation}
e^{-i\tilde\lambda_it}=e^{i\theta}\text{sign}(a_i)    \label{eqn:transfer_cond}
\end{equation}
equality can be achieved for $F_{\text{max}}$. To specify the eigenvalues $\tilde\lambda_i$, one selects a time $t\gg\pi/\Delta'$ to be the state transfer time, where $\Delta'=\min_{i\neq j}|\lambda_i-\lambda_j|$. Each of the $\lambda_i$ is truncated to an accuracy of $\pi/t$ such that the number of intervals $k$ of $\pi/t$ that separate $\tilde\lambda_i$ and $\tilde\lambda_j$ determines the value of $e^{-i(\tilde\lambda_i-\tilde\lambda_j)t}=(-1)^k$. Each eigenvalue can then be shifted by no more than 1 interval to ensure that Eqn.~(\ref{eqn:transfer_cond}) is satisfied. Finally, for symmetric effective Hamiltonians, an inverse eigenvalue problem (IEP) \cite{Hoc74} can be solved to match the new eigenvalues, and perfect state transfer is assured \footnote{IEPs can also be solved in the non-symmetric case when supplied with the desired values of $a_n$. While the existence of solutions is not always guaranteed, as it is in the symmetric case, the perturbative approach means that by starting close to an existing solution, a solution is expected. $F_{\text{max}}$ is no longer guaranteed to be 1, but is $1/\text{poly}(N)$ in the absence of string tension.}. The linearity of the problem \cite{Hoc74} guarantees that the $J_i$ are only shifted by a small amount, $O(1/(\delta t))$, from their initial values and hence the Hamiltonian perturbations are still precisely that -- perturbations satisfying $\|h_i\|\leq 1$. For example, using the Toric code with $J_i=\half$, $\Delta'\sim\delta/N^2$, and the logical rotation occurs in a time $\sim N^2/\delta$.

Having designed the Hamiltonian perturbation assuming that an error occurs on a specific qubit, it is worth observing that this is not the limit of what can be achieved. For example, the construction only acts on a single row of the lattice, and similar terms can be implemented on each row (the terms from one row commute with those on other rows). Furthermore, if an error were to occur on a different qubit in the row, then \cite{kay-2006b} reveals that this is propagated to a single error on a different spin in the state transfer time. This is proved by using a duality mapping to the perfect state transfer chain \cite{christandl}, which implements the perfect mirroring of states in all excitation subspaces \cite{Christandl:2004a}, due to the applicability of the Jordan-Wigner transformation. Denoting the controlled-NOT gate, with control $i$ and target $j$, by $C^i_j$, the mapping can be given as
$$
V\delta HV^{\dagger}=\half\delta\sum_iJ_i(X_{2i,0}X_{2i+2,0}+Y_{2i,0}Y_{2i+2,0})
$$
where
$$
V=\left(\prod_{i=0}^{N-2}\!\!C^{(2i+1,-1)}_{(2i,0)}C^{(2i+1,1)}_{(2i,0)}\right)\!\!\left(\prod_{i=1}^{N-2}\!\!C^{(2i,0)}_{(2i-2,0)}\right).
$$
The order of the second product is important, because the controlled-NOTs do not commute; the right-most term should be $C^{(2,0)}_{(0,0)}$ (i.e.~this is the first gate to be applied) and $i$ increases to the left. The first excitation subspace of the transformed Hamiltonian maps to $\{U_i\ket{\psi}\}$, and the second excitation subspace maps to $\{U_iU_j\ket{\psi}\}_{i>j}$. Hence, any single error at a site $(2i,0)$ with $i>0$ is described by $U_{i}U_{i-1}\ket{\psi}$ and transfers to $U_{N-i-1}U_{N-i-2}\ket{\psi}$ in the perfect state transfer time. During the evolution between these two states, there is some probability that it has gone through an intermediate state from which the stored information cannot be reliably recovered.

{\em Self Correction:} The result on the Toric code is applicable in a wide range of cases, since the IEP technique allows almost any effective Hamiltonian to be recast into a perfect transfer system (provided that a single error can be connected by a polynomial sequence of local operations to a state that gets falsely corrected, as is certainly the case for stabilizer Hamiltonians), giving a high likelihood of destruction of the stored data. This statement is predicated on the assumption that the minimum gap between eigenvalues, $\Delta'$, is not exponentially small. We shall now analyze the 2D Ising model, the archetypal construction of a self-correcting code for classical data, to see how it coincides with these conditions. Note that proving that the 2D Ising model is protected against this class of noise does not constitute a proof that the system is self correcting. Rather, the proof can only be used to show that certain systems are not self-correcting.

The Ising Hamiltonian is defined as
$$
H_I=-\half\sum_{\langle i,j\rangle}Z_iZ_j
$$
where the sum is over nearest neighbour pairs, $\langle i,j\rangle$, of an $N\times N$ square lattice with periodic boundary conditions and qubits are placed on the vertices. The two ground states $\ket{0}^{\otimes N^2}$ and $\ket{1}^{\otimes N^2}$ are capable of storing classical information (a single $Z$ error destroys phase information). If a pattern of $X$ errors affects one of the states, the resultant state is an eigenvector of $H_I$, and has an energy equal to the surface area of the pattern. To convert between the two ground states requires a pattern of $X$ rotations on every single qubit. One can consider writing this as a progressive sequence, starting from a single $X$ error, adding one rotation at each step, progressively filling up entire rows (this sequence provides a convenient labelling  of the qubits, 1 to $N^2$). All other choices of sequence have a similar behaviour. Since the stabilizers of this Hamiltonian are different to those of the Toric code, $\delta H$ requires revision:
$$
\delta H=\half\delta\sum_{i=1}^{N^2-2}J_iX_{i+1}(\identity-Z_{i}Z_{i+2})+\half\delta\sum_iB_i(\identity-Z_iZ_{i+1}).
$$
Motivation for the design is identical, and coincides with \cite{kay-2006b}. To ease the analysis, without affecting the eventual outcome, $\delta H$ can be further modified -- in the terms which would introduce an $X$ which would complete an entire row ($X_{Ni}$ for integer $i$), we instead apply $X_{Ni}X_{Ni+1}$, which also starts the next row, keeping the surface area of the block of $X$s constant. Taking $J_i=1,B_i=0$, the effective Hamiltonian (which is an $M\times M$ matrix, where $M=N(N-1)-2$) can be written as
\begin{eqnarray}
&&H_{\text{eff}}=(N+1)\identity+\delta\sum_{i=1}^{M-1}(\ket{i}\bra{i+1}+\ket{i+1}\bra{i})   \nonumber\\
&&-\sum_{i=1}^{N-1}2(N-i)(\proj{i}+\proj{M+1-i}). \nonumber
\end{eqnarray}
The first step is to determine the eigenvalues. Since $\delta$ is small, perturbation theory can be applied to the initial eigenvalues of $2(i+1)$ for $i=1\ldots N$. All of these are repeated twice, except for $2(N+1)$, which is repeated $(N-1)(N-2)-2$ times. Degenerate perturbation theory is first applied to the repeated $2(N+1)$ energy, which splits the degeneracy at first order,
$$
E_i=2(N+1)+2\delta\cos\left(\frac{i\pi}{(N-1)(N-2)-1}\right)
$$
for integer $1\leq i\leq(N-1)(N-2)-2$. Now, considering all the other energies $2(i+1)$, by symmetry, all possible shifts for the repeated eigenvalues are identical, until the $(M+1-2i)^{th}$ order. So, for $i=1$, the energy difference between the two originally degenerate levels is $O(\delta^{N^2})$, and the strategy of fixing the eigenvalues would require an exponentially long evolution, which is not prohibitive. In fact, no set of quasi-local perturbations can give destruction of the data; no Hamiltonian term acting locally on $k$ qubits can change a state $U_i\ket{\psi}$ to any other state $U_j\ket{\psi}$ for $|i-j|>k$, so the effective Hamiltonian is limited to being $k$-banded, i.e.~only the first $k$ sub-diagonals can be non-zero. Furthermore, these are all of strength $\leq\delta$. Thus, the perturbation calculation of eigenvalue splittings applies equally to all possible perturbations. By identical arguments, one can similarly prove that the 4D Toric code \cite{DKLP02a} is stable for storage of quantum information with respect to this noise model.

The critical feature of the 2D Ising model is the fact that the energies of the sequence of $X$ operations changes with the length of the sequence, scaling with $N$, which is the property known as string tension. This is essential because our technique has to be able to protect against all local errors, not just the single error that has been considered so far. In order to achieve this, it is necessary that when considering (degenerate) perturbation theory on the effective Hamiltonian, in order to connect any pair of initial and final states, the number of intermediate steps must grow with $N$ and each sequence of steps between unperturbed eigenstates must go through a number of distinct energy states that grows with $N$, necessitating an increasing order of perturbation theory. Finally, in order for the perturbation to remain a perturbation, the smallest gap between unperturbed eigenvalues must be constant. So, the spectrum for the string must vary with length, and the maximum energy must grow with $N$.

{\em In Conclusion,} we have given an explicit error mechanism that shows that the 2D Toric code is not self-correcting i.e.~that stored data survives for no longer than $O(N)$, where there are $N^2$ qubits in the system \footnote{ A stronger error model that one could consider is the introduction of errors at a fixed density i.e.~as the system size increases, so does the number of errors. In that scenario, the perturbations (through a trivial modification of $\delta H$) would only have to propagate the errors over constant sized blocks, and the upper bound to the survival time is reduced to $O(1)$.}. This mechanism, in which Hamiltonian perturbations play a leading role, if not the precise scaling, is capable of destroying the information stored in many systems that do not exhibit string tension, motivating the expectation that string tension is a necessary condition for self-correcting quantum memories. In contrast to previous approaches, we have been able to consider a dynamical process that does not rely on equilibrium properties, but describes how long storage is possible for on approach to equilibrium. Note, however, that our treatment has little to do with the thermalization process itself due to the limited consideration made of the bath interaction, as indicated by the high purity preserved in our model.

The interplay between noise and Hamiltonian perturbations is a feature that may prove important in a number of other settings. For example, in topological quantum computation, a non-Abelian model is defined by a Hamiltonian on a 2D lattice, and the idea is to store and compute using the excitations of the model by braiding them around each other. A fault occurs due to noise if a pair of anyons is created by some noise event and braids around one of the computational anyons. Unlike the case of memories, in topological quantum computation, we have the ability to regularly perform active error correction. The general expectation is that the fault-tolerant threshold should be similar to that for the circuit model of computation, but that the inherent robustness of the gates should make it easier to get the gate fidelities over the required threshold. Since perturbations in the Hamiltonian can provide the propagation mechanism to convert localized faults into genuine gate errors in the topological model, this implies that in schemes without string tension between the anyon pairs, the expected robustness of the braiding operations is reduced.

{\em Acknowledgments:} The author wishes to thank J.~I.~Cirac for useful discussions. This work is supported by DFG (FOR 635 and SFB 631), EU (SCALA), Clare College, Cambridge and the National Research Foundation \& Ministry of Education, Singapore.

\end{document}